\documentstyle[preprint,tighten,aps]{revtex}       
\begin{document}
\draft

\title{On the use of a non-extensive distribution in the solar plasma}

\author{A. Lavagno and P. Quarati} 
\address{Dipartimento di Fisica, Politecnico di Torino, C.so Duca degli 
Abruzzi 24, I-10129 Torino, Italy}
\address{Istituto Nazionale di Fisica Nucleare, 
Sezioni di Torino e di Cagliari}

\maketitle

\begin{abstract}
We describe the physical motivations why we are urged to look for an ionic 
equilibrium distribution function slightly different from the Maxwellian to 
calculate nuclear reaction rates in the solar core plasma answering, 
at the same time, to a recent paper by Bahcall et al. (astro-ph/0010055).
\end{abstract}

\vspace{1cm} 

In a recent paper, Bahcall et al. \cite{bah} discuss motivations why Salpeter 
formula for screening of nuclear reactions in the Sun is correct (in contrast 
with conclusions contained in several other papers) and show five different 
derivations of the Salpeter screening factor $f_0$.

A paper of us \cite{npb}
is also included in this critical discussion (although our work 
is not concerned directly on the factorizable screening factor $f_0$ but,  
rather, on the evaluation of the reaction rates) because of the 
use we make of an ionic equilibrium distribution which differs very slightly 
from the Maxwellian distribution. 
The reason of using this equilibrium distribution is related to the need of 
including many-body effects. 
It was already proposed {\it ad hoc} by Clayton \cite{clay} 
and we can see, a posteriori, that coincides to the equilibrium 
distribution of the non-extensive thermostatistics developed by 
Tsallis \cite{tsa} 
and used recently in many different physical applications \cite{tsaweb}, 
as for instance,  
gravitational \cite{gala} and high energy problems \cite{albe}. 

In our calculations the screening factor $f_0$ is taken to be unity because 
its effect on the reaction rates is negligible 
if compared to the important depletions (or enhancements) of 
the standard rates due to other effects (electromagnetic fluctuations in 
a plasma) responsible of the non-extensive equilibrium distribution function. 
Also, by following the derivation of $f_0$ in Ref.\cite{bah2} and using 
non-extensive distribution, we can calculate for the solar core about 
the same value of the screening factor. For comments based on Molecular 
Dynamics Model (MDM) on the validity of Salpeter screening formula see 
Ref. \cite{shaviv}. Several authors are actually working on 
MDM within non-extensive Tsallis statistics \cite{tsa2}. 
It seems to us 
obvious that, if all the derivations of the Salpeter formula are based on 
the validity of the assumptions needed for the Boltzmann-Gibbs statistics, 
the equilibrium distribution must be, of course, the Maxwellian distribution. 
However the motivations that deviations from these assumptions are not correct 
should be explicitly given. 

Results within non-extensive statistics go toward a better agreement 
with the experimental results of the calculated neutrino fluxes. 
In spite of this achievement, we never excluded the possibility of neutrino 
oscillation or new physics because only experiments will give the 
correct answer on this problem. 
At the same time, we have also shown that modifi\-ca\-tions 
of the reaction rates 
do not affect bulk properties of the gravitationally stabilized solar core 
such as sound speed or hydrostatic equilibrium, depending on mean values 
obtained averaging over Maxwellian distribution function \cite{inno}. 

We confirm that Tsallis distribution is an equilibrium distribution and that 
the argument that there is enough time to arrive at equilibrium after a 
reaction is fully accomplished by the Tsallis equilibrium distribution 
that can be derived in, 
at least, three different ways containing not alternative reasons. 

In one approach 
we exploit the knowledge that the distribution in the
solar interior cannot be too much different from the Maxwellian one
and add small corrections (higher-order terms in a derivative expansion)
to the coefficients of the standard Fokker-Plank equation. 
Tsallis' distribution is immediately generated \cite{kania97}.

A second approach focuses on the electric microfields that have been
shown to exists in the solar plasma: the time-spatial fluctuations in 
the particles positions produce specific fluctuations of microscopic 
electric field in a given point of the plasma.  
We have shown that such electric microfields in the solar plasma  
(with energy density of the order of $10^{-16}$ MeV/fm$^3$) 
implies a deviation from the Maxwellian distribution; the 
entity depending on the value of the plasma parameter and 
on the correlation among ions \cite{npb,ala}.

The third approach has been just started and aims to connect the distribution
of collective variables 
to memory effects and long-time correlations
between velocities \cite{valuev98,cora}. 
There should exist solutions compatible with
the Tsallis' distribution and/or other non-Maxwellian distributions.

These three approaches are not exhaustive and not necessary alternative.
Nevertheless, it is suggestive that all of them point in the same direction:
{\em the Maxwell-Boltzmann distribution of velocity should have small but
nonnegligible corrections in the solar plasma and the Tsallis' distribution
could provide a better description}.

Normal stellar matter, such as the one in the Sun, is non-degenerate,
{\em i.e.}, quantum effects are small (in fact, they are small for electrons
and completely negligible for ions), it is non-relativistic, and it is in
good thermodynamical equilibrium. On this ground, the particle velocity
distribution is almost universally taken to be 
a Maxwell-Boltzmann distribution, without much questioning.

However, derivations of the ubiquitous Maxwell-Boltzmann distribution are
based on several assumptions~\cite{kania97}.
In a kinetical approach, one assumes (1)
that the collision time be much smaller than the mean time between collisions,
(2) that the interaction be sufficiently local, (3) that the velocities of
two particles at the same point are not correlated (Boltzmann's
Stosszahlansatz), and (4) that energy is locally conserved when using only the
degrees of freedom of the colliding particles (no significant amount of
energy is transferred to collective variables and fields). In the 
equilibrium statistical mechanics approach, one uses the assumption
that the velocity probabilities of different particles are independent,
corresponding to (3), and that the
total energy of the system could be expressed as the sum of a term quadratic
in the momentum of the particle and independent of the other variables,
and a term independent of momentum, but if (1) and (2) are not valid
the resulting effective two-body
interaction is not local and depends on the momentum and energy of
the particles.
Finally, even when the one-particle distribution is Maxwellian,
additional assumptions about correlations between particles are necessary
to deduce that the relative-velocity distribution, which is the
relevant quantity for rate calculations, is also Maxwellian.

At least in one limit the Maxwell-Boltzmann distribution can be rigorously 
derived: systems that are
dilute in the appropriate variables, whose residual interaction is
small compared to the one-body energies. In spite of the fact
that the effects of the residual interaction cannot be neglected,
as a good first approximation the solar interior can be studied in this dilute
limit; therefore, it is reasonable to suppose that the velocity distribution
in the Sun is not too far from the Maxwellian one.

Likewise, one often assumes that the solar core
could be treated as an ideal (Debye) plasma. However, there are physical
conditions and/or specific applications that needs higher accuracy for
which becomes necessary to take into account modifications of the
standard plasma theory.

When $\Gamma \approx 0.05 \div 0.1$,  
the mean Coulomb energy potential is not much
smaller of the thermal kinetic energy and the screening length 
$R_D\approx a$. It is not possible to clearly separate individual and
collective degrees of freedom. The presence of at least two different
scales of energies of the same rough size produces deviations from the
standard statistics which describe the system in terms of a single
scale, $kT$.

The reaction time necessary to build up screening after a hard collision can
be estimated from the inverse solar plasma frequency 
$t_{pl}=\omega_{pl}^{-1} = 
\sqrt{m/ (4\pi n e^2 )} \approx 10^{-17}$~sec, 
and it is comparable to the
collision time $t_{coll} = \langle\sigma v n \rangle^{-1} \approx 10^{-17}$ 
~sec. Therefore, several collisions are likely necessary before the particle
looses memory of the initial state and the scattering
process can not be considered Markovian. In addition, screening
starts to become dynamical: the time necessary to build up again the
screening after hard collisions is not negligible any more.

Many remarks we have just made are reported in well known books on 
statistical mechanics, kinetic and plasma theory and in works by 
astrophysicist as, for instance, Paquette et al. \cite{paque} and 
Cox et al. \cite{cox}. 
Let us report, specifically, from Cox et al.:  
{\it ``one usually assumes (1) negligible radiative forces,  
(2) complete ionization, (3) Maxwellian velocity distributions and the 
same kinetic temperature for all ions and electrons, (4) diffusion velocities 
which are much smaller than mean thermal velocities, (5) no magnetic fields, 
(6) collisions dominated by ``classical'' interactions between two point 
particles, and (7) plasma which can be considered a diluite gas, i.e., the 
ideal gas equation of state applies. [..] When the number of particles in a 
Debye sphere around an ion is not $\gg 1$, assumption (6) and (7) above are 
inappropriate. Ions are no longer effectively screened from their 
surrounding, and multiple particle collisions and collective effects 
become important. 
The ``classical'' approach to calculating transport properties of ions using 
Boltzmann's equation becomes invalid. In the Sun, at number for particles 
per Debye sphere is only about 1.4 at the base of the convection zone, 
and 2.2 at the core, so the ``diluite gas'' approximation fails.''}

Finally, let us recall that the introduced equilibrium distribution 
contains the factor $e^{-(1-q)/2 (E/kT)^2}$ where $q$ 
is the Tsallis parameter. 
In the solar core plasma we have found for it the value 
$\vert (1-q)/2\vert\approx 0.003$ ($q=1$ corresponds to the Boltzmann-Gibbs 
statistics and Maxwellian distribution). 
In spite of the very small deviation of the behavior at high energy of the 
Tsallis distribution respect to the standard curve 
the effect on the reaction 
rates (depletion or enhancement) is quite important \cite{npb,kania97,ala}. 

{}

\end{document}